\documentclass[12pt]{article}

\usepackage{scicite}
\usepackage{times}
\usepackage{graphicx}
\usepackage[htt]{hyphenat}

\usepackage{graphicx,color,inputenc,bm,amsmath}
\usepackage{caption}
\usepackage{amssymb}

\newenvironment{sciabstract}{%
\begin{quote} \bf}
{\end{quote}}

\title{Quantifying team chemistry in scientific collaboration} 

\author
{Gangmin Son$^{1}$, Jinhyuk Yun$^{2\ast}$, Hawoong Jeong$^{1,3\ast}$\\
\\
\normalsize{$^{1}$Department of Physics, Korea Advanced Institute of Science and Technology (KAIST)}\\
\normalsize{$^{2}$School of AI Convergence, Soongsil University}\\
\normalsize{$^{3}$Center for Complex Systems, KAIST}\\
\\
\normalsize{$^\ast$To whom correspondence should be addressed;}\\
\normalsize{E-mail: jinhyuk.yun@ssu.ac.kr or hjeong@kaist.edu.}
\\
}

\date{}

\begin{document} 

\baselineskip24pt

\maketitle 

\begin{sciabstract}

Team chemistry is the holy grail of understanding collaborative human behavior, yet its quantitative understanding remains inconclusive. To reveal the presence and mechanisms of team chemistry in scientific collaboration, we reconstruct the publication histories of 560,689 individual scientists and 1,026,196 duos of scientists. We identify ability discrepancies between teams and their members, enabling us to evaluate team chemistry in a way that is robust against prior experience of collaboration and inherent randomness. Furthermore, our network analysis uncovers a nontrivial modular structure that allows us to predict team chemistry between scientists who have never collaborated before. Research interest is the highest correlated ingredient of team chemistry among six personal characteristics that have been commonly attributed as the keys to successful collaboration, yet the diversity of the characteristics cannot completely explain team chemistry. Our results may lead to unlocking the hidden potential of collaboration by the matching of well-paired scientists.

\end{sciabstract}

\section*{Introduction}
Teams are ubiquitous and essential in various human activities, posing the intriguing subject of building a great team~\cite{mcewan_teamwork_2014,kozlowski_enhancing_2006,wuchty_increasing_2007,guimera_team_2005,woolley_evidence_2010}. One of the challenges in this goal is that a good team should result in more than the sum of its constituents. For example, assembling the top players on a football team does not guarantee the best performance~\cite{swaab_too-much-talent_2014}. The concept of team chemistry, therefore, has been debated to explain these discrepancies in diverse fields such as sports and business~\cite{ryan_intangibles_2020,belbin_management_1983}. In addition, team performance can be affected by chance and shared experience~\cite{aoki_luck_2017,mukherjee_prior_2019}. To characterize team chemistry, it is thus critical to understand these effects.

From this perspective, science is an appropriate domain to study the concept in a sophisticated way. Burgeoning digitization of academic data has recently spawned a new field of science, namely the science of science~\cite{fortunato_science_2018,zeng_science_2017}, that has shed light on a variety of aspects of knowledge creation and consumption. Furthermore, one of the most fundamental elements of modern science is the increasing dominance of teams~\cite{wuchty_increasing_2007}. Although team formation procedures and their linkages to scientific impact have been explored~\cite{petersen_quantifying_2015,guimera_team_2005,wu_large_2019}, little is known about whether the impact of a team's papers is distinct from that of its members working separately.

Scientific impact is usually measured by the number of citations a work receives~\cite{wang_quantifying_2013}. Recent studies on the scientific impact patterns of individual scientists have revealed that the random-impact rule and the so-called $Q$-model can untangle the role of productivity, luck, and a scientist's own ability, or $Q$, within her or his career~\cite{sinatra_quantifying_2016,janosov_success_2020}. Beyond individual careers, scientific collaboration is often repeated enough to constitute a \textit{team career}~\cite{petersen_quantifying_2015}, raising the following questions: Is the $Q$-model also in charge of the dynamics of team careers? Can a team's $Q$ be explained only by the combination of its members' $Q$s? If not, what are the mechanisms and origins of the discrepancy? Is it possible to foresee the discrepancy before the collaboration? Answers to these questions would certainly motivate individual scientists that seek well-suited partners. Moreover, such insight may also help institutions and policymakers better manage human resources to accelerate scientific innovation on a global scale.

Here, we reveal quantitative aspects of the nebulous concept of team chemistry in scientific collaboration. We use a large-scale bibliometric data set to reconstruct the publication histories of 560,689 individual scientists and 1,026,196 duos of scientists as the most elementary form of a team. We find that the random-impact rule governs the evolution of both individual and collaborative scientific impact, which allows us to untangle the role of longevity, luck, and a team's (and also an individual scientist's) own ability that characterize a career. Building on this finding, we identify ability discrepancies, which are not blurred by inherent randomness in scientific careers, between two actual scientists working together and separately, supporting the presence of team chemistry in scientific collaboration.

Furthermore, we explore the mechanisms of team chemistry using a network approach. We detect a nontrivial modular structure in team chemistry networks, which opens the door to predicting the interaction type between two scientists who have never collaborated. The prediction task does not require any additional information except for the interaction types of other duos in the network, implying that team chemistry originates from the appropriate combinations of the intrinsic features of individual scientists, the patterns of which are encoded by the network. Finally, we collect six classes of individual attributes---gender, ethnicity, academic age, research interests, affiliations, and working countries---and show that research interest is the most relevant to team chemistry, although its (dis)similarity does not have a consistent relation with team chemistry across disciplines.

\section*{Results}

\subsection*{Understanding dynamics of team careers in science}

\begin{figure}
    \centering
    \includegraphics[width=.7\textwidth]{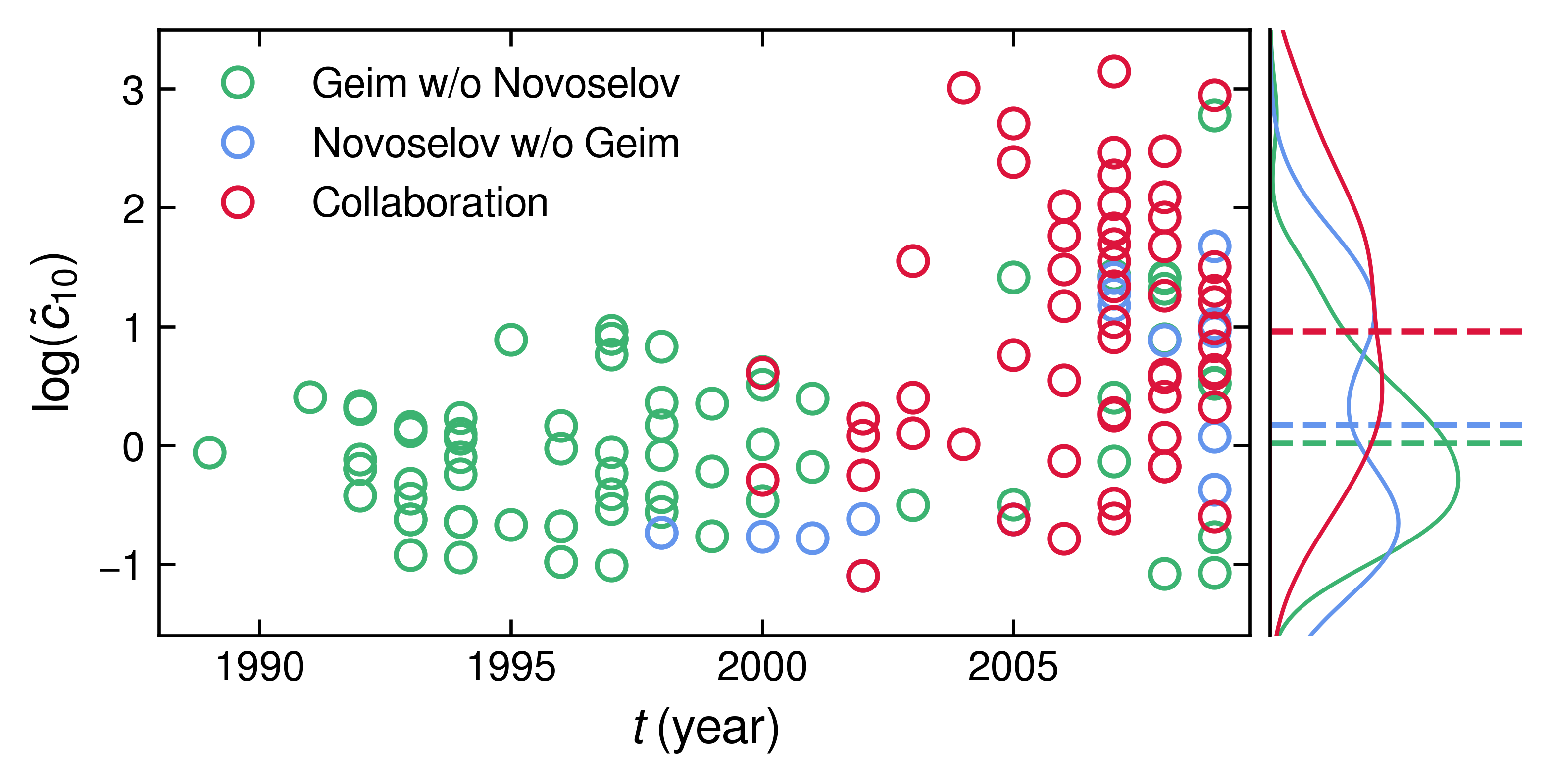}
    \caption{\textbf{Team career.}
    The publication histories of Andre Geim and Konstantin Novoselov, who shared the 2010 Nobel Prize in Physics. Each point represents a research paper written by Geim (green), Novoselov (blue), and both of them together (red). We measured the impact of a paper, $\tilde{c}_{10}$, using the normalized citation count 10 years after its publication (see Materials and Methods), and present the distribution of each type of paper on the right along with their geometric means (dashed horizontal lines). At first glance, Geim and Novoselov have tended to produce more impactful publications when they collaborate than when they work separately.}
    \label{fig:team_career}
\end{figure}

We begin by gathering the publication histories of individual scientists and teams from a variety of disciplines (see section~\ref{subsection:Data description}). Each paper's scientific impact is evaluated by $\tilde{c}_{10}$, which is the normalized number of citations received 10 years after publication (see section~\ref{subsection:Scientific impact}). Figure~\ref{fig:team_career} depicts the collaborative publication history, or team career, of Dr. Andre Geim and Dr. Konstantin Novoselov, who were jointly awarded the 2010 Nobel Prize in Physics, as well as their individual publication histories. At first look, they appear to publish more influential papers when they work together, implying that they have synergy. Individual careers, though, are known to be affected by random fluctuations \cite{sinatra_quantifying_2016}, and thus we must similarly comprehend the dynamics of team careers to determine whether they truly have synergy. 

\begin{figure}
    \centering
    \includegraphics[width=\textwidth]{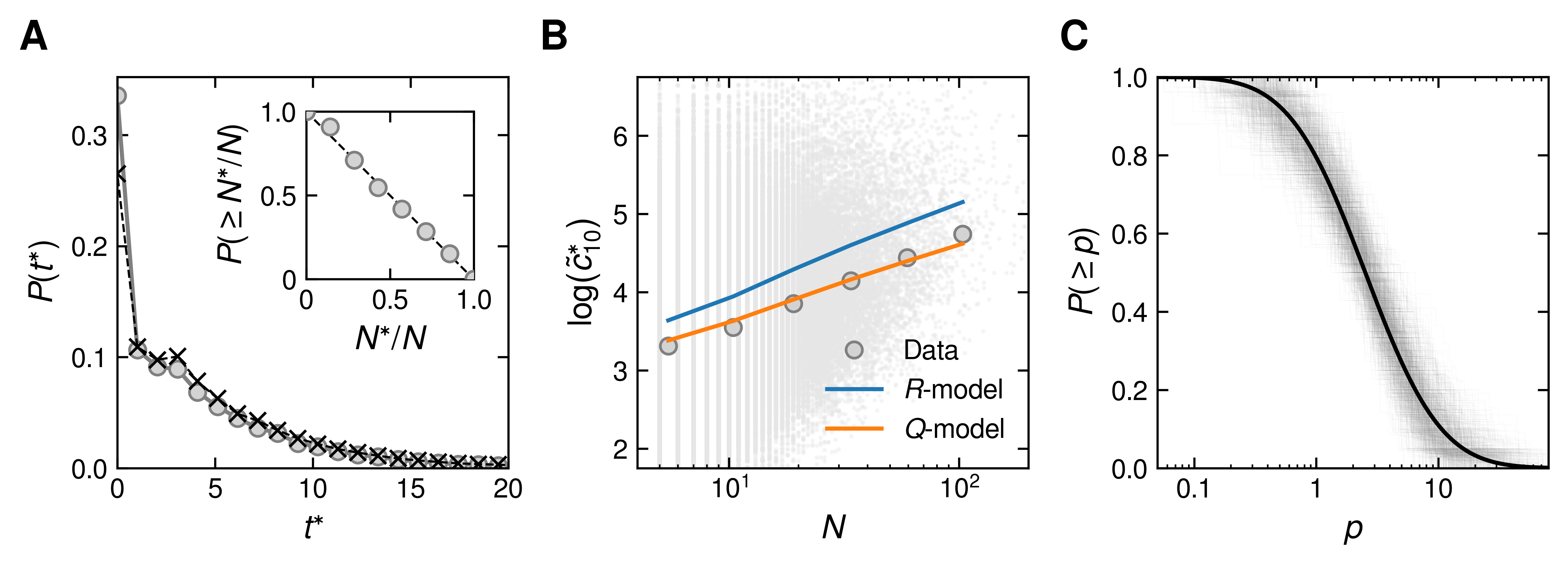}
    \caption{\textbf{Impact patterns in team careers.}
    (\textbf{A}) The distribution of the timing of the highest-impact paper $P(t^{*})$ and the cumulative distribution of the position $P(\leq N^{*}/N)$, indicating that the highest-impact paper is distributed randomly in a team career. (\textbf{B}) $\langle \log{\left(\tilde{c}_{10}\right)} \rangle$ as a function of $N$. Each point in the scatter plot corresponds to a duo of scientists, and the gray circles represent the log-binned mean of the data. The prediction of the $Q$-model (orange) agrees well with the data, in contrast to the R-model (blue). (\textbf{C}) The collapse of the cumulative distributions $P(\geq\tilde{c}_{10}/Q)$. Each gray curve corresponds to a duo, and the black curve represents the universal distribution $P(p)$.
    }
    \label{fig:career_pattern}
\end{figure}

First, we study when a team publishes its most influential paper by measuring $P(t^{*})$ and $P(N^{*}/N)$, which are the distributions of the timing $t^{*}$ and the relative position $N^{*}/N$ of the highest-impact article in a sequence of $N$ publications, to see if the random-impact rule holds for team careers. We discover that the highest-impact paper is distributed randomly in team careers (Fig.~\ref{fig:career_pattern} and Fig.~\ref{fig:Qval-team-A}), implying that the random-impact rule applies to both individual and team careers (Fig.~\ref{fig:Qval-ind-A}). This study shows that collaborators' shared experience does not boost their chances of success in a systematic way; that is, collaborating with an old partner does not ensure more success than collaborating with a new partner, and vice versa.

However, we cannot explain the increase in a team's highest impact $\tilde{c}_{10}^{*}$ as a function of the number of collaborations between its members $N$ when we assume a null $R$-model in which each team has the same impact distribution $P(\tilde{c}_{10}^{*})$ (orange line in Fig.~\ref{fig:career_pattern}B). As a result, their impact distributions differ, which can be explained by the $Q$-model \cite{sinatra_quantifying_2016}. This model states that the impact of an article $\alpha$ published by scientist $i$ is determined by a multiplicative process as follows:

\begin{equation}
    \tilde{c}_{10, i\alpha} = Q_{i} p_{\alpha},
\end{equation}

\noindent where $Q_i$ is an intrinsic variable of an individual or team $i$, and $p_\alpha$ is a stochastic variable, usually interpreted as the potential of paper $\alpha$ or luck. To obtain the values of $Q_i$ and $p_\alpha$, we estimate the mean vector $\mu$ and the covariance matrix $\Sigma$ of the underlying joint probability $P(p,Q,N)$ by using a maximum-likelihood approach (see section~\ref{subsection:parameter_estimate}).

We calculate $\log{\tilde{c}_{10}^{*}}$ as a function of $N$ based on the estimated parameters. The $Q$-model can predict the increase of the highest impact in a career, as seen in Fig.~\ref{fig:career_pattern}B. Furthermore, the negligible mixed terms of $\Sigma$, $\sigma_{pN}^2 = \sigma_{pQ}^2 \approx 0$, suggest that $p_\alpha$ is independent of $Q_i$ and $N_i$ (table~\ref{table:params}); that is, luck is scientist-independent and universal. As a result, all teams' $P(\tilde{c}_{10}/Q)$ distributions collapse into a single curve $P(p)$, resulting in a remarkably similar pattern across the teams. Our findings are consistent across disciplines (figs.~\ref{fig:Qval-team-B} and~\ref{fig:Qval-team-C}) as well as individual careers (figs.~\ref{fig:Qval-ind-B} and~\ref{fig:Qval-ind-C}). In summary, we show that the $Q$-model, which was originally designed to understand the evolution of individual success, also drives team success patterns.

\subsection*{Sum of its parts, luck, or chemistry?}

As we observed in the previous section, not only individuals but even duos of scientists have their own unique metric $Q$, which represents their ability to attract citations. This raises an intriguing question of how the team's ability corresponds to its members' abilities. It is reasonable to assume that the higher the average individual abilities, the greater the team's ability. Accordingly, we attempt to anticipate $Q_{\{i,j\}}$ of the collaboration between two scientists $i$ and $j$ from the generalized mean of the individual members' $Q$ without each other, $Q_{i-j}$ and $Q_{j-i}$, as follows:

\begin{equation}
\label{eqn:generalized mean}
    Q_{\{i,j\}}^{\rm{expect}} = \left(\frac{Q_{i-j}^\beta + Q_{j-i}^\beta}{2}\right)^{1/\beta}.
\end{equation}

\begin{figure}
    \centering
    \includegraphics[width=.7\textwidth]{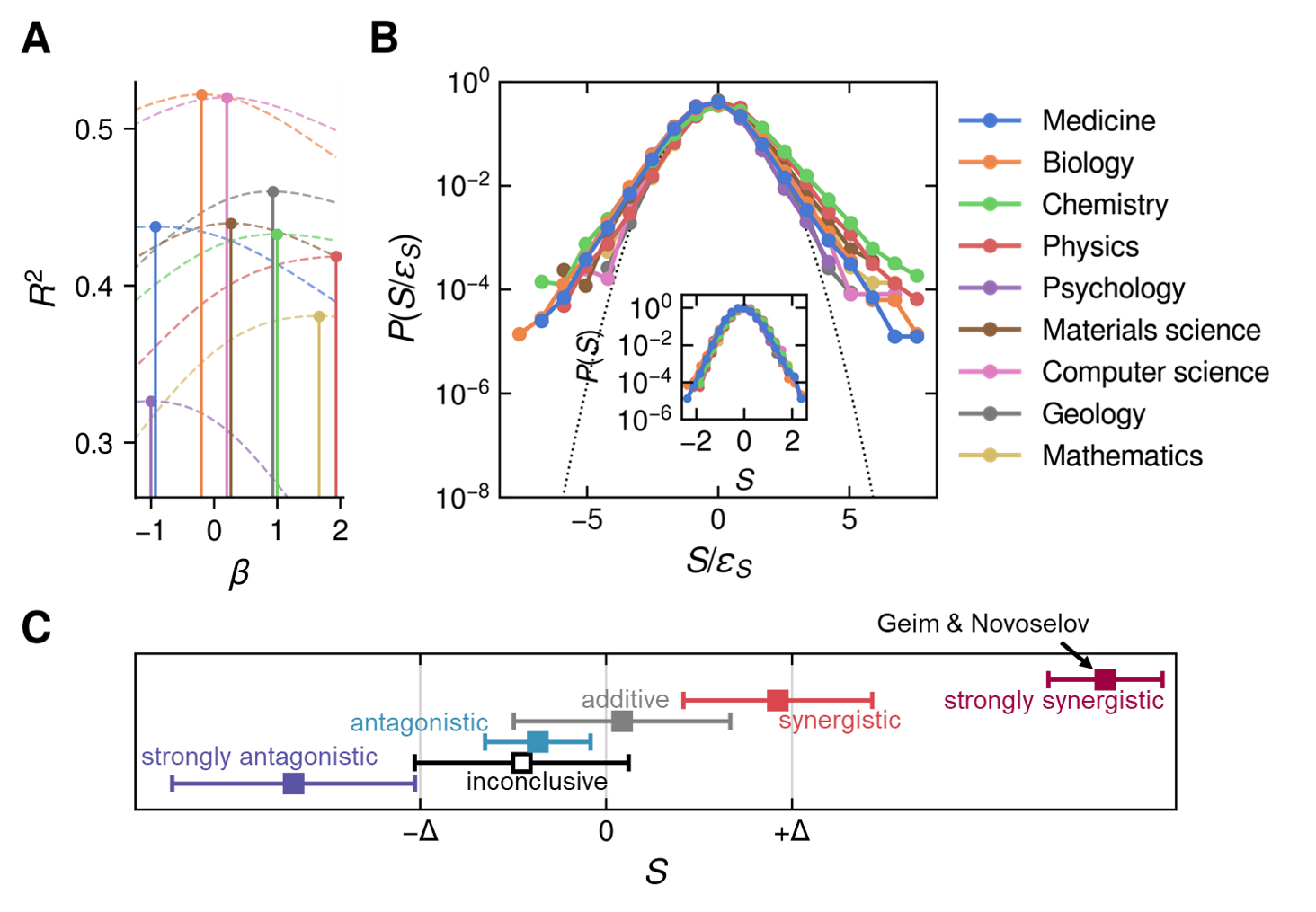}
    \caption{
    \textbf{Quantification of team chemistry.}
    (\textbf{A}) The coefficient of determination $R^{2}$ between $Q^{\rm{measure}}$ and $Q^{\rm{expect}}$ as a function of the generalized mean parameter $\beta$. We infer the optimal additivity relation which maximizes $R^{2}$ for each discipline. (\textbf{B}) The normalized distributions of $S/\epsilon_S$ for each discipline. The dotted line represents the standard normal distribution, and the inset shows the distributions of $S$. (\textbf{C}) Classification of team chemistry. The error bars correspond to 95\% confidence intervals calculated from the $Q$-model. As an illustrative example, the team chemistry between Geim and Novoselov is strongly synergistic.
    }
    \label{fig:chemistry}
\end{figure}

\noindent The additivity rule for each discipline is then defined by estimating $\beta$, which maximizes the coefficient of determination $R^2$ between $Q^{\rm{expect}}$ and the measured $Q$, $Q^{\rm{measure}}$. For the extreme cases, $\beta = \infty$ yields the maximum among the given values, whereas $\beta = -\infty$ yields the minimum among the given values; thus, larger $\beta$ values indicate that the team $Q$ depends on its higher-$Q$ members, and vice versa. Here, we find that the optimal $\beta$ differs among fields (Fig.~\ref{fig:chemistry}, $\beta$ ranges from $-1.00$ to $1.93$). For instance, the higher-$Q$ member is the most crucial for greater scientific impact in physics, with a $\beta$ of $1.93$.

The moderate values of $R^2$ (0.31--0.51) in Fig.~\ref{fig:chemistry}A imply significant deviations from the optimum additivity rules. We use $S = \log (Q^{\rm{measure}}/Q^{\rm{expect}})$ to calculate the deviation and find that the distribution $P(S)$ has a finite width (the standard deviation ranges 0.37--0.43, see inset of Fig.~\ref{fig:chemistry}B). How can we account for these discrepancies? The limited number of publications in a career implies an inherent uncertainty $\epsilon_Q$ in the measurement of $Q$ because the progress of a scientific career is driven by a stochastic process, as we observed in the previous sections. Even if a team follows the additivity rule precisely, a small number of publications of certain teams or individuals can give the impression that the team is breaking the rule. Therefore, we normalize $S$ by its uncertainty $\epsilon_S$, which is derived using the error propagation rule along with the $\epsilon_Q$s of a team and its members (see section~\ref{subsection:uncertainty_S}). The distribution of $S/\epsilon_{S}$ would follow the probability density function of the standard normal distribution $\mathcal{N}(0,1)$ if there is no team chemistry in scientific collaboration and all teams follow the obtained additivity criterion. Surprisingly, a disparity is observed in the derived $P(S/\epsilon_S)$ (Fig.~\ref{fig:chemistry}B; kurtosis ranges 0.06--1.83). As a result, the $Q$-model's inherent uncertainty is insufficient to explain the observed non-additive effects.

Can the disparity shown in Fig.~\ref{fig:chemistry}B now be regarded as team chemistry? The fact that repeat collaboration can involve more than two scientists is one possible cause of these deviations. In other words, the chemistry between two scientists could be a result of additional co-authors. We calculated their credit share \cite{shen_collective_2014} and subtracted it from the team careers to remedy this situation. We discovered that large discrepancies in ability persist after the credit allocation and change little (see section~\ref{section:credit_allocation} and fig.~\ref{fig:credit_allocation_team_chemistry}). In this way, our results support the existence of team chemistry in scientific collaboration.

Consequently, we can classify teams into three major categories: synergistic, additive, and antagonistic (Fig.~\ref{fig:chemistry}C). Note that if two scientists have a small number of publications, we consider their type of team chemistry inconclusive since the uncertainty of $S$ is too high. Our classification results demonstrate that synergy, defined as team outcomes greater than the sum of its parts, is not universal; around half of the teams are additive, while some are even antagonistic (fig.~\ref{fig:chemistry_distribution}).

\subsection*{A network approach to explore the mechanisms of team chemistry}

\begin{figure}
    \centering
    \includegraphics[width=.7\textwidth]{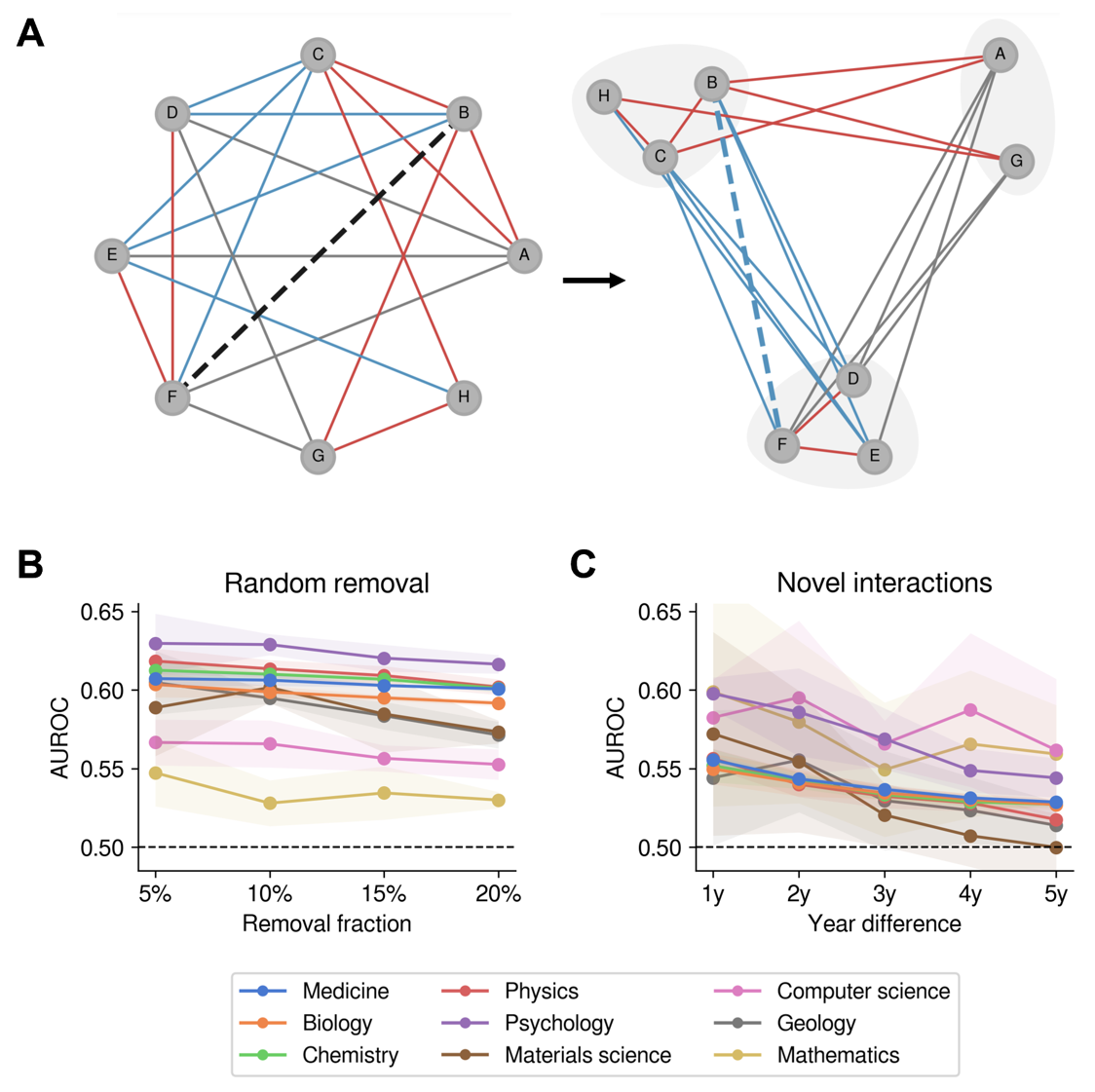}
    \caption{\textbf{Modular structure in team chemistry networks and its predictability}
    (\textbf{A}) The diagram illustrates a hypothetical team chemistry network. We hypothesize that the interaction type of a scientist pair is determined by the combination of their attributes. In such a case, we can infer the block structure from the known interactions and estimate the types of unknown interactions. (\textbf{B} and \textbf{C}) We measure the predictability by the area under the receiver operating characteristic curve (AUROC).  (\textbf{B}) shows the AUROC for random removal as a function of removal fraction, and (\textbf{C}) displays the AUROC for novel interaction prediction as a function of the difference between the observation year that we use to detect the modular structure and the predicted year. The shaded areas correspond to the 95\% confidence interval for each discipline.}
    \label{fig:sbm}
\end{figure}

What distinguishes some teams as synergistic and others as non-synergistic? One potential scenario is that team chemistry is determined by a combination of individual characteristics such as skill set or personality. Here, we adopt a network approach to explore the generative mechanisms of team chemistry. Consider a network of team chemistry where nodes represent individual scientists and edges reflect the observations of pairwise team chemistry, with the edge types representing the three defined categories of team chemistry (Fig.~\ref{fig:sbm}A). According to the above scenario, the edge type between any two nodes is defined as a function of certain characteristics of the two scientists. In the hypothetical situation in Fig.~\ref{fig:sbm}A, the team chemistry between two groups \{B,C,H\} and \{D,E,F\} emerges as antagonistic. Although the type of team chemistry between two scientists, B and F, is unknown, their group memberships can be inferred by using the other edges, and thus their type of team chemistry can be predicted as antagonistic.

These group-dependent interactions can be mathematically formalized by stochastic block models (SBMs)~\cite{holland_stochastic_1983,newman_structure_2016}. If the generative mechanisms of team chemistry follow the scenario in Fig.~\ref{fig:sbm}A, the unobserved types of team chemistry can be predicted by statistical inference of the SBM. To test this, we construct a network of the obtained team chemistry types for each discipline (table~\ref{table:SSIN}). We infer the block membership of each scientist (node) in a Bayesian framework~\cite{peixoto_bayesian_2019,guimera_missing_2009,guimera_network_2013}, and predict the unknown interactions based on these inferred block partitions~(see Materials and Methods). Specifically, we perform two different tasks: (i) predicting random-removal interactions, and (ii) predicting emerging interactions in a series of yearly snapshots. The predictive accuracy of each task is measured by the area under the receiver operating characteristic curve (AUC). Notably, without any information but the interaction types of other teams, our approach is able to predict unknown types of team chemistry (Fig.~\ref{fig:sbm}B and~C), although the small margins ($\rm{AUC}<0.65$) suggest that external factors may also affect the emergence of team chemistry.

We further investigate whether the obtained block patterns correlate with six types of individual attributes: gender, ethnicity, academic age, research interests, affiliations, and working countries (see section~\ref{section:metadata}). We measure the correlation between the probability that two scientists were grouped into the same block and the similarity of their individual attributes for each attribute type, the values of which are higher as the attributes are more associated with the generative mechanisms of the block structure (see Materials and Methods). We find no attribute strongly associated with the block structure (Pearson's $r<0.1$; table~\ref{table:SBM_metadata}). However, we cannot entirely deny the possibility that these attributes are related to team chemistry because the structural pattern captured by metadata and that captured by an SBM can differ, as mentioned in~\cite{peel_ground_2017}. Therefore, we should examine the relationships between the attributes and team chemistry regardless of the block structure.

\begin{figure}
    \centering
    \includegraphics[width=.7\textwidth]{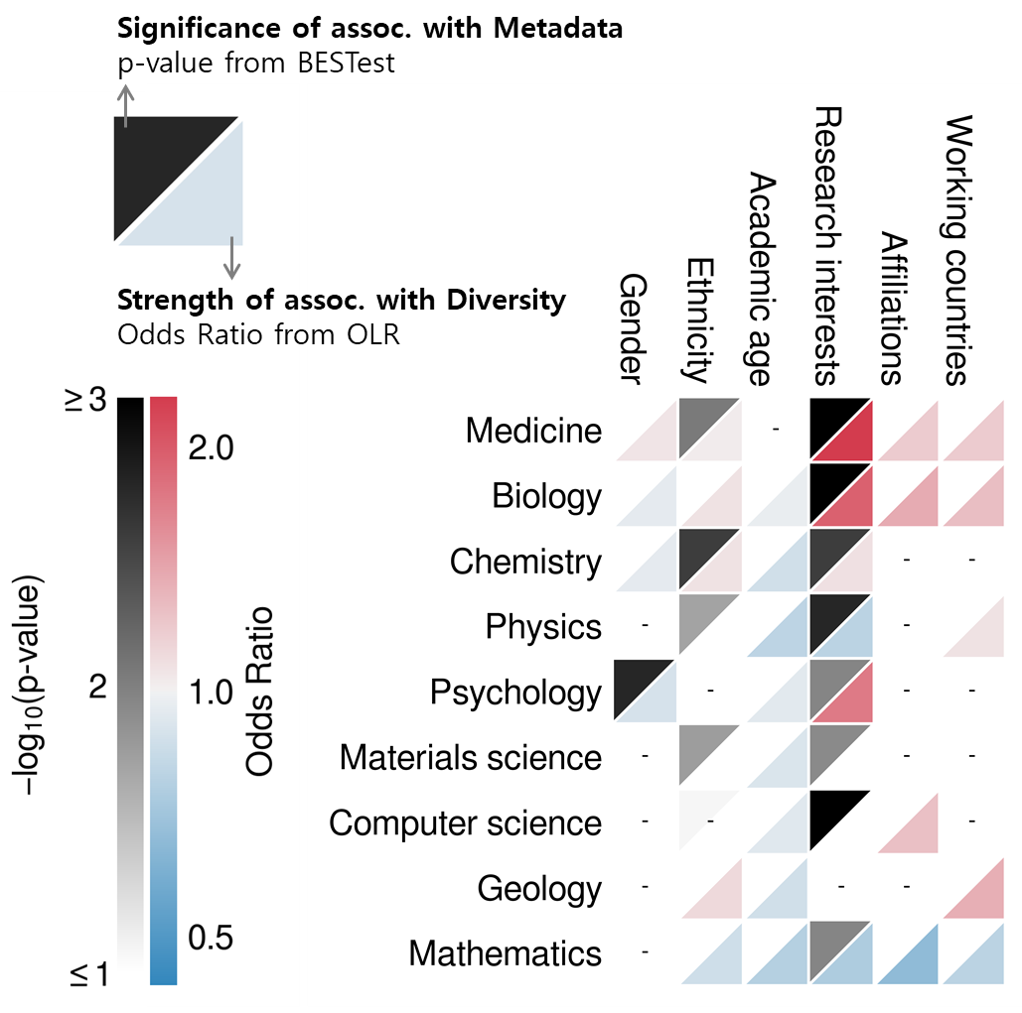}
    \caption{\textbf{Correlations between individual attributes and team chemistry.} 
    The results of two statistical analyses describe the relationships between six individual attributes and team chemistry across disciplines. The upper wedge corresponds to the results of the BESTest, and the lower wedge corresponds to the results of ordinal logistic regression with team chemistry as the ordinal outcome and the diversity of each attribute as the explanatory variable (see Materials and Methods).}
    \label{fig:metadata}
\end{figure}

We perform the inspection by measuring the significance of the association between each attribute type and team chemistry using the Blockmodel Entropy Significance Test (BESTest; see Materials and Methods) in~\cite{peel_ground_2017}. As shown in Figure~\ref{fig:metadata} (upper wedges), while attributes exhibiting significant association are rare, research interests is the attribute most closely related to team chemistry for most fields ($P<0.012$; except geology, $P=0.112$). Our next interest is interpreting the associations in the view of their diversity, which is occasionally attributed as the key to successful collaboration~\cite{alshebli_preeminence_2018,nielsen_opinion_2017,adams_fourth_2013,jones_multi-university_2008,freeman_collaborating_2015,uzzi_atypical_2013}. We examine the connection between the diversity of each attribute and team chemistry by obtaining the odds ratio from ordinal logistic regression. An odds ratio above 1 indicates that an increase in diversity leads to a more synergistic type of team chemistry, while a value less than 1 implies a negative effect of diversity. Figure~\ref{fig:metadata} (lower wedges) shows that the strengths of association are weak in general ($|\log_{2}{\rm{Odds Ratio}}|<1$; except research interests in medicine with $1.20$). Remarkably, we find no robust patterns across the cases of research interests; in some fields, diversity in this attribute leads to synergy, while in other fields it can suppress synergy. These inconsistent directions of association allow us to conclude that team chemistry might be driven by more complicated mechanisms, rather than by simple factors such as attribute diversity or homogeneity as previous studies have pointed out~\cite{alshebli_preeminence_2018,nielsen_opinion_2017,adams_fourth_2013,jones_multi-university_2008,freeman_collaborating_2015,uzzi_atypical_2013}.

\section*{Discussion}

In this study, we investigate the previously vague concept of team chemistry in a quantitative and mechanistic way. Our main contribution is twofold. First, we identify discrepancies between the performances achieved by two scientists working together and separately that cannot be explained by the effects of career longevity, productivity, or the inherent randomness in their scientific careers, supporting the presence of team chemistry in scientific collaboration. Second, we find that team chemistry networks have nontrivial modular structures, allowing us to predict unknown interactions and investigate the associations between individual attributes and team chemistry.

It is noteworthy that the random-impact rule also holds for team careers. Unlike in team sports \cite{mukherjee_prior_2019}, prior shared experience does not improve the probability of success in scientific collaboration. However, simultaneously, a correlation between collaboration longevity and performance has been observed \cite{petersen_quantifying_2015}, raising a question: Do two scientists keep their ties because they get satisfactory achievements, or do their long-lasting ties lead to success? Our results suggest that the former scenario might be more plausible, yet we leave the question for further study at this moment.

The network analysis used in this paper provides a new framework for understanding how collaborative scientific impact emerges and the role of individual attributes. It may yield insights into $Q$, namely how its non-additive effect results in team chemistry. For example, it is still debated whether citations or $Q$ indicate scientific quality more than the other~\cite{aksnes_citations_2019,simkin_success_2020}. If $Q$ is more related to the ability to produce high-quality research, more content-related factors such as research interests and skill sets would correlate with team chemistry. Although our results highlighted the correlation between team chemistry and individual research interests across different fields, it is a promising future direction to collect additional individual attributes in a more sophisticated way and to investigate their correlations with team chemistry using our approach. Because scientific impact is a multifaceted concept~\cite{aksnes_citations_2019}, team chemistry in terms of other dimensions, such as disruptiveness~\cite{wu_large_2019}, is also an interesting issue.

Team chemistry among more than two scientists is also an important topic. Although we here focused on pairwise interactions, scientific collaboration intrinsically involves higher-order interaction~\cite{battiston_networks_2020,battiston_physics_2021}. Considering team chemistry between more than two scientists, however, encounters difficulties. For instance, the longevity of repeat collaboration decays rapidly as the team size increases (fig.~\ref{fig:why_pairwise}), amplifying the uncertainty in the measurement of $Q$. In addition, higher-order team chemistry requires higher-order additivity. In other words, it is necessary to measure and rule out the effects of each sub-team. For example, to quantify team chemistry among three scientists, measuring the team chemistry of each pairwise interaction is necessary, yet such could be impossible because of the sparsity of the bibliometric data. In addition, a more complicated approach is necessary to define the additivity rule for higher-order interactions. These problems could be tackled by advanced methods such as higher-order interaction reconstruction \cite{young_hypergraph_2021} or well-controlled social experiments \cite{clauset_data-driven_2017}.

Our findings may have implications for not only the scientific community but also policymakers. The basic motivation of scientific team building is to improve outcomes from collaboration. For this purpose, extending beyond previous studies that merely predicted future collaboration~\cite{kong_exploring_2017,schleyer_conceptualizing_2012,araki_interdisciplinary_2017}, our work paves the way to the matching of underrepresented scientists with unrealized potential, which could accelerate scientific innovation on a global scale. As a final remark, we note the need for caution in exploiting predictability in order not to overlook the unpredictable nature of science~\cite{clauset_data-driven_2017}.

\section*{Materials and Methods}

\subsection*{Data}

We use the Microsoft Academic Graph (MAG)~\cite{sinha_overview_2015,wang_microsoft_2020} released in October 2019, which contains 228,996,078 articles and 231,970,249 authors. The data set is distributed as multiple tables in \texttt{TSV} format. We identified the authors of a paper using the \texttt{AuthorId} element in the \texttt{PaperAuthorAffiliations} table and adopted an additional process of author disambiguation and conflation to reconstruct the publication histories of 560,689 individual scientists and 1,026,196 duos of scientists spanning 9 fields of study. The scientific impact of a paper is calculated using rescaled citations received 10 years after its publication, $\tilde{c}_{10}$, from the \texttt{PaperReferences} table. The Supplementary Materials contain more detailed information.

\subsection*{$Q$-model}

The $Q$-model~\cite{sinatra_quantifying_2016} is a mechanistic model that generates impact sequences based on the random-impact rule, which states that impact is distributed randomly during a career. In this model, a stochastic process $c_{i\alpha}=Q_{i} p_{\alpha}$ determines the impact $c_{i\alpha}$ of a paper $\alpha$ produced by scientist $i$, where $p_\alpha$ corresponds to inherent randomness. The underlying assumption is that the length of the sequence and $Q_i$ are unaffected by $p_{\alpha}$. To validate this, we calculate the model parameters and see if the correlations between $p$ and $N$ and between $p$ and $Q$ are negligible (table~\ref{table:params}).

\subsection*{Stochastic Block Model}

The stochastic block model (SBM) \cite{holland_stochastic_1983, newman_structure_2016} is a generative network model in which adjacency matrix $\bm{A}$ is generated with probability $P(\bm{A}|\bm{b})$, where $\bm{b}$ is a group membership vector with entries $b_i\in\{1,...,B\}$. The posterior block partition distribution is written as
\begin{equation}
    P(\bm{b}|\bm{A}) = \frac{P(\bm{A}|\bm{b})P(\bm{b})}{P(\bm{A})}.
\end{equation}
Based on this, we can infer the block structure of the team chemistry networks. Specifically, we use a nonparametric Bayesian method \cite{peixoto_nonparametric_2017, peixoto_bayesian_2019} implemented in the Python library \texttt{graph-tool} (https://graph-tool.skewed.de).

Team chemistry networks possess edge types $\bm{x}={x_{ij}}$, where $x_{ij} \in \{\text{synergistic}, \text{additive}, \text{antagonistic}\}$. Our purpose is not to predict whether an edge exists between two nodes but rather to predict the type of edge if one exists. To infer the unknown edge types $\bm{\delta x}$, we sample partitions from the posterior and average the probability of the unknown interactions as follows:
\begin{align}
    P(\bm{\delta x}|\bm{A^{O}},\bm{x^{O}})
    &= \sum_{\bm{b}}{P(\bm{\delta x}|\bm{b})P(\bm{b}|\bm{A^{O}},\bm{x^{O}})}\nonumber\\
    &= \frac{ \sum_{\bm{b}}{P(\bm{\delta x}|\bm{b})P(\bm{A^{O}},\bm{x^{O}}|\bm{b})P(\bm{b})} }
    { \sum_{\bm{b}}P(\bm{A^{O}},\bm{x^{O}}|\bm{b})P(\bm{b}) },
\end{align}
where $\bm{A^O}$ and $\bm{x^O}$ denote the observed edges and their types. Based on this, we calculate the probability of the edge type for each unknown interaction. Then, we measure the prediction performance as AUROC.

\subsection*{Blockmodel Entropy Significance Test}

The Blockmodel Entropy Significance Test (BESTest) \cite{peel_ground_2017} is a statistical test for determining the significance of the association between a network structure and metadata. The statistic of the BESTest is

\begin{equation}
    P\ \text{value} = \text{Pr}[\mathcal{S}^{\prime} \leq \mathcal{S}],
\end{equation}

\noindent where $\mathcal{S}$ is the entropy of the SBM with consideration of the metadata as a partition, and $\mathcal{S}^{\prime}$ is the same as $\mathcal{S}$ but using a random permuted version of the metadata. In detail, when it comes to the attributes that are assigned multiple elements per node (research interests, affiliations, and working countries), we use the mixed-membership SBM~\cite{godoy-lorite_accurate_2016}. The Supplementary Materials contain more detailed information.

\subsection*{Diversity}

We estimate the diversity of each of the six traits we gathered to observe the impact of diversity on collaboration. When it comes to gender and ethnicity, $1-\delta_{a_i a_j}$ is used to measure diversity, with $i$'s attribute $a_i$ being a categorical variable. For the academic age groups, the difference in years is used to reflect diversity. Then for research interests, affiliations, and working countries, we define diversity as $1 - J(a_i, a_j)$, where $i$'s attribute $a_i$ is a set of traits for each author and $J$ is the Jaccard similarity.

We employ ordinal logistic regression to investigate the association between team chemistry and the diversity of attributes. The ordinal outcome is team chemistry, and the explanatory variable is the diversity. We use the log odds ratios to measure the intensity and direction of an association.

\section*{Acknowledgement}
The National Research Foundation (NRF) of Korea funded by the Korean government supported this work through Grant No. NRF-2017R1A2B3006930 (G.S., H.J.) and NRF-2020R1A2C1100489 (J.Y.). The Korea Institute of Science and Technology Information (KISTI) also supported this work by providing KREONET, the high-speed internet connection. The funders had no role in the study design, data collection and analysis, decision to publish, or preparation of the manuscript.

\section*{Author contributions}
All three authors designed the research and wrote the paper. Gangmin Son collected and analyzed the data. 

\section*{Competing interests}
The authors declare that they have no competing interests.

\pagebreak

\clearpage

\end{document}